\documentclass[pdftex,twocolumn,epjc3]{svjour3}          

\RequirePackage[T1]{fontenc}

\smartqed  
\RequirePackage{graphicx}
\RequirePackage{mathptmx}      
\RequirePackage{flushend}
\RequirePackage[numbers,sort&compress]{natbib}
\RequirePackage[colorlinks,citecolor=blue,urlcolor=blue,linkcolor=blue]{hyperref}

\journalname{Eur. Phys. J. C}

\begin{document}

\title{Coupled dark energy: a dynamical analysis with complex scalar field}


\author{Ricardo C. G. Landim\thanksref{e1,addr1}
}

\thankstext{e1}{rlandim@if.usp.br}

\institute{Instituto de F\'isica, Universidade de S\~ao Paulo\\
 Caixa Postal 66318,  05314-970 S\~ao Paulo, S\~ao Paulo, Brazil\label{addr1}
}

\date{Received: date / Accepted: date}

\maketitle

\begin{abstract}
The dynamical analysis for coupled dark energy with dark matter is presented, where a complex scalar field is taken into account and it is considered in the presence of a barothropic fluid. We consider three dark energy candidates: quintessence, phantom and tachyon. The critical points are found and their stabilities analyzed, leading to the three cosmological
eras (radiation, matter and dark energy), for a generic
potential. The results presented here enlarge the previous analyses found in the literature. \end{abstract}

\section{Introduction}

Observations of Type IA Supernova indicates that the universe undergoes an accelerated expansion \cite{reiss1998, perlmutter1999}, which is dominant at present times ($\sim$ 68\%) \cite{Planck2013cosmological}. Despite of ordinary matter, the remaining $27\%$ is an unknown form of matter that interacts in principle only gravitationally, known as dark matter. The nature of the dark sector is still mysterious and it is one of the biggest challenges in the modern cosmology. The simplest dark energy candidate is the cosmological constant, whose equation of state $w_\Lambda=p_\Lambda/\rho_{\Lambda}=-1$ is in agreement with the Planck results \cite{Planck2013cosmological}. This attempt, however,  suffers from the so-called cosmological constant problem, a huge discrepancy of 120 orders of magnitude between the theoretical prediction and the observed data. 

Among a wide range of alternatives, a scalar field is a viable candidate to be used. Its usage includes the canonical scalar field, called ``quintessence'' \cite{peebles1988,ratra1988,Frieman1992,Frieman1995,Caldwell:1997ii},  and the scalar field with the opposite-sign in the kinetic term, known as ``phantom'' \cite{Caldwell:1999ew,Caldwell:2003vq}. Beyond the real scalar field case, a complex quintessence was also used in ref.~\cite{Gu2001} to account the acceleration of the universe. The $U(1)$ symmetry associated with this complex scalar leads to a more sofisticated structure for the dark sector, and unless the stardard model of particle physics is a very special case in the nature, there is no reason (apart from simplicity) not to consider a richer physics of the dark sector\footnote{ A current example of a vector field that perhaps interacts with dark matter is the so-called ``dark photon'' (see \cite{Essig:2013lka} for a quick review).}.

Another possibility of noncanonical scalar field is the tachyon and it comes from string theory. In the bosonic string, its ground state is the tachyon field, whereas in supersymmetric string theory a real tachyon is present in non-BPS Dp-branes, while the complex tachyon appears in a brane-anti-brane system \cite{Sen:2004nf}.  The tachyon potential has a minimum \cite{Sen:1998sm, Sen:1999xm} and at this minimum  the tachyon field behaves like a pressureless gas \cite{Sen:2002in}. As soon as tachyon condensation in string theory had been proposed, tachyon was also regarded as a dark energy candidate \cite{Padmanabhan:2002cp,Bagla:2002yn,Abramo:2003cp}. 

Still regarding the dynamical dark energy, there exists the possibility of interaction between dark energy and dark matter \cite{Wetterich:1994bg,Amendola:1999er}, since their densities are comparable and, depending on the coupling used, the interaction can alleviate the coincidence problem \cite{Zimdahl:2001ar,Chimento:2003iea}. This approach was applied to phantom and tachyon as well, in refs. \cite{Guo:2004vg,Cai:2004dk,Guo:2004xx,Bi:2004ns,Gumjudpai:2005ry,micheletti2009}.

When a scalar field is in the presence of a barothropic fluid (with equation of state $w_m=p_m/\rho_m$)  the relevant evolution equations can be converted   into an autonomous system and the asymptotic states of the cosmological models can be analyzed. Such approach was done for uncoupled dark energy (quintessence, tachyon field and phantom field  for instance \cite{copeland1998,ng2001,Copeland:2004hq,Zhai2005,DeSantiago:2012nk}) and coupled dark energy \cite{Amendola:1999er,Gumjudpai:2005ry,TsujikawaGeneral,amendola2006challenges,ChenPhantom,Mahata:2015lja,Landim:2015poa,Khurshudyan:2015yca,Khurshudyan:2015mva}.  Since the complex scalar field and the coupled dark energy are generalizations of the real field and the uncloupled case, respectively, we aim to study both possibilities together, in the light of the linear dynamical systems theory. Thus, we investigate in this paper the critical points that come from the evolution equations for the complex scalar field (quintessence, phantom and tachyon), considering the possibility of interaction between the two components of the dark sector. The dynamical equations are derived and the critical point are analyzed, to find out what kind of universe can come up. As we have said, this is a natural extension of the previous works \cite{Amendola:1999er,Gumjudpai:2005ry,ChenPhantom,Landim:2015poa}. As a result, we have found no new fixed points, however there are some crucial differences if compared with the case of real field.

The rest of the paper is organized as follows. In section \ref{de} we present the basics of the interacting dark energy and the dynamical analysis theory, where we show the couplings we have used. In sections \ref{quintphantom} and \ref{tachdyn} we use the dynamical system theory for the canonical (and phantom) and the tachyon field, respectively. Within the respective section we show the critical points with their stabilities, and also the viable  sequence of cosmological eras (radiation-matter-dark energy). Section \ref{conclu} is reserved for conclusions. We use Planck units ($\hbar=c=1 =M_{pl}=1$) throughout the text.

\section{Interacting dark energy and dynamical analysis}\label{de}

We  consider that dark energy is described by a scalar field with energy density $\rho_\phi$ and pressure $p_\phi$, and with an equation of state  given by $w_\phi=p_\phi/\rho_\phi$. We assume that the scalar field is coupled with dark matter, in such a way that total energy-momentum is still conserved. In the flat Friedmann-Robertson-Walker background with a scale factor $a$, the continuity equations for both components and for radiation are

\begin{equation}\label{contide}
\dot{\rho_\phi}+3H(\rho_\phi+p_\phi)=-\mathcal{Q},
\end{equation}

\begin{equation}\label{contimatter}
\dot{\rho_m}+3H\rho_m=\mathcal{Q},
\end{equation}

\begin{equation}\label{contirad}
\dot{\rho_r}+4H\rho_r=0,
\end{equation}

\noindent respectively, where $H=\dot{a}/a$ is the Hubble rate,  $\mathcal{Q}$ is the coupling and the dot is a derivative with respect to the cosmic time $t$. The indices $m$ and $r$ stand for matter and radiation, respectively. The case of $\mathcal{Q}>0$ corresponds to dark energy transformation into dark matter, while $\mathcal{Q}<0$ is the transformation in the opposite direction. In principle, the coupling  can depend on several variables $\mathcal{Q}=\mathcal{Q}(\rho_m,\rho_\phi, \dot{\phi},H,t,\dots)$, so we assume for the canonical scalar (quintessence) and for the phantom field the coupling is $\mathcal{Q}=Q \rho_m\dot{\phi}$ \cite{Wetterich:1994bg,Amendola:1999er}, while for the tachyon field the coupling is $\mathcal{Q}=Q \rho_m\rho_\phi\dot{\phi}/H$ \cite{Landim:2015poa}, where $Q$ is a positive constant. The case with negative $Q$ is similar and we will not consider it here because the minus sign of the case $\mathcal{Q}<0$ can be absorved into the $\dot{\phi}$, instead of considering $Q<0$.

To deal with the dynamics of the system, we will define dimensionless variables. The new variables are going to characterize a system of differential equations in the form

\begin{equation}
X'=f[X],
\end{equation}

\noindent where $X$ is a column vector of dimensionless variables and the prime is the derivative  with respect to $ \log a$, where we set the present scale factor $a_0$ to be one. The critical points $X_c$ are those ones that satify $X'=0$. In order to study stability of the fixed points, we consider linear perturbations $U$ around them, thus $X=X_c+U$. At the critical point the perturbations $U$ satisfy the following equation

\begin{equation}
U'=\mathcal{J}U,
\end{equation}

\noindent where $\mathcal{J}$ is the Jacobian matrix. The eigenvalues of $\mathcal{J}$ determine if the critical points are stable (if all eigenvalues are negative), unstable (if all eigenvalues are positive) or saddle points (if at least one eigenvalue is positive and the others are negative, or vice-versa).

\section{Quintessence and phantom dynamics}\label{quintphantom}

The complex scalar field $\Phi$ can be written as $\Phi=\phi e^{i\theta}$, where $\phi$ is the absolute value of the scalar field and $\theta$ is a phase. Both canonical and phantom fields are described by the  Lagrangian

\begin{equation}\label{scalar}
 \mathcal{L}=-\sqrt{-g}\left(\frac{\epsilon}{2}\partial^\mu\Phi\partial_\mu\Phi+V(|\Phi|)\right),
\end{equation} 

\noindent where $V(|\Phi|)$ is the potential for the complex scalar and we consider it depends only on the absolute value  of the scalar field $\phi\equiv|\Phi|$. We have $\epsilon=+1$ for the canonical field (quintessence) and $\epsilon=-1$ for the phantom field. For a homogeneous field $\phi\equiv\phi(t)$ and $\theta\equiv\theta(t)$, in an expanding universe with Friedmann-Robertson-Walker metric with scale factor $a\equiv a(t)$, the equations of motion are

\begin{equation}\label{eqmotionscalar1}
 \epsilon \ddot{\phi}+3\epsilon H\dot{\phi}+V'(\phi)-\epsilon \phi \dot{\theta}^2=0,
\end{equation}

\begin{equation}\label{eqmotionscalar2}
 \epsilon \ddot{\theta}+\left(3 H+\frac{2\dot{\phi}}{\phi}\right)\dot{\theta}=0.
\end{equation} 

\noindent where the prime denotes derivative with respect to $\phi$. In the uncloupled case Eq. (\ref{eqmotionscalar2}) gives rise to the effective potential in (\ref{eqmotionscalar1}), which is $\frac{d}{d\phi}\left(\frac{\omega^2}{2a^6}\frac{1}{\phi^2}+V(\phi)\right)$ \cite{Gu2001}, where $\omega$ is an integration constant interpreted as angular velocity. The first term in the brackets drives $\phi$ away  from zero  and the factor $a^{-6}$ may make the term decreases very fast, provided that $\phi$ does not decrease faster than $a^{-3/2}$.  

We assume the interaction between the scalar field with dark matter through the coupling  $Q \rho_m\dot{\phi}$ and it enters in the right-hand side of Eq. (\ref{eqmotionscalar1}).

In the presence of matter and radiation, the Friedmann equations for the canonical (phantom) field are

\begin{equation}\label{eq:1stFEmatterS}
  H^2=\frac{1}{3}\left(\frac{\epsilon}{2}\dot{\phi}^2+\frac{\epsilon}{2}\phi^2\dot{\theta}^2+V(\phi)+ \rho_m+\rho_r\right),
\end{equation}

\begin{equation}\label{eq:2ndFEmatterS}
  \dot{H}=-\frac{1}{2}\left(\epsilon\dot{\phi}^2+\epsilon\phi^2\dot{\theta}^2+\rho_m+\frac{4}{3}\rho_r\right),
\end{equation}

\noindent and the equation of state becomes

\begin{equation}\label{eqstateS}
 w_\phi=\frac{p_\phi}{\rho_\phi}=\frac{\dot{\phi}^2+\phi^2\dot{\theta}^2-2\epsilon V(\phi)}{\dot{\phi}^2+\phi^2\dot{\theta}^2+2\epsilon V(\phi)}.
\end{equation}

We are now ready to proceed the dynamical analysis of the system.

\subsection{Autonomous system} 

The dimensionless variables are defined as

\begin{eqnarray}\label{eq:dimensionlessXYS}
 x_1\equiv  &\frac{\dot{\phi}}{\sqrt{6}H}, \quad x_2\equiv \frac{\phi\dot{\theta}}{\sqrt{6}H}, \quad x_3\equiv \frac{\sqrt{6}}{\phi}, \quad y\equiv \frac{\sqrt{V(\phi)}}{\sqrt{3}H},\nonumber\\
  & z\equiv \frac{\sqrt{\rho_r}}{\sqrt{3}H}, \quad \lambda\equiv -\frac{V'}{V}, \quad \Gamma\equiv \frac{VV''}{V'^2}.
\end{eqnarray}

The dark energy density parameter is written in terms of these new variables as

\begin{equation}\label{eq:densityparameterXYS}
 \Omega_\phi \equiv \frac{\rho_\phi}{3H^2} =\epsilon x_1^2+\epsilon x_2^2+y^2,
 \end{equation}

\noindent so that Eq. (\ref{eq:1stFEmatterS}) can be written as 

\begin{equation}\label{eq:SomaOmegasS}
\Omega_\phi+\Omega_m+\Omega_r=1,
\end{equation}

\noindent where the matter and radiation density parameter are defined by $\Omega_i=\rho_i/(3H^2)$, with $i=m,r$. From Eqs. (\ref{eq:densityparameterXYS}) and (\ref{eq:SomaOmegasS}) we have that $x_1$, $x_2$ and $y$ are restricted in the phase plane by the relation

\begin{equation}\label{restrictionS}
0\leq \epsilon x_1^2+\epsilon x_2^2+y^2\leq 1,
 \end{equation}
 
\noindent due to $0\leq \Omega_\phi\leq 1$. Notice that if $y=0$ the restriction (\ref{restrictionS}) forbids the possibility of phantom field ($\epsilon=-1$) because for this case $\Omega_\phi<0$.

The equation of state $w_\phi$  becomes

\begin{equation}\label{eq:equationStateXYS}
 w_\phi =\frac{\epsilon x_1^2+\epsilon x_2^2-y^2}{\epsilon x_1^2+\epsilon x_2^2+y^2},
\end{equation}

\noindent which is a trivial extension of the real scalar field case. The total effective equation of state is

\begin{equation}\label{eq:weffS}
 w_{eff} = \frac{p_\phi+p_r}{\rho_\phi+\rho_m+\rho_r}=\epsilon x_1^2+\epsilon x_2^2-y^2+\frac{z^2}{3},
\end{equation}

\noindent with an accelerated expansion for  $w_{eff} < -1/3$.  The dynamical system for the variables  $x_1$, $x_2$,  $x_3$, $y$, $z$  and $\lambda$ are

\begin{eqnarray}\label{dynsystemS}\label{eq:dx1/dnS}
\frac{dx_1}{dN}&=-3x_1+x_2^2x_3+\frac{\sqrt{6}}{2}\epsilon y^2\lambda-\frac{\sqrt{6}}{2}\epsilon Q(1-x_1^2-x_2^2\nonumber\\ &-y^2-z^2)
-x_1H^{-1}\frac{dH}{dN},\end{eqnarray}

\begin{equation}\label{eq:dx2/dnS}
\frac{dx_2}{dN}=-3x_2-x_1x_2x_3-x_2H^{-1}\frac{dH}{dN},\end{equation}

\begin{equation}\label{eq:dx3/dnS}
\frac{dx_3}{dN}=-x_1x_3^2,\end{equation}

\begin{equation}\label{eq:dy/dnS}
\frac{dy}{dN}=-\frac{\sqrt{6}}{2}x_1 y\lambda-yH^{-1}\frac{dH}{dN},
\end{equation}

\begin{equation}\label{eq:dz/dnS}
\frac{dz}{dN}=-2z-zH^{-1}\frac{dH}{dN},
\end{equation}

\begin{equation}\label{eq:dlambda/dnS}
\frac{d\lambda}{dN}=-\sqrt{6}\lambda^2 x_1\left(\Gamma-1\right),
\end{equation}

\noindent where

\begin{equation}\label{}
H^{-1}\frac{dH}{dN}=-\frac{3}{2}(1+\epsilon x_1^2+\epsilon x_2^2-y^2)-\frac{z^2}{2}.
\end{equation}

\subsection{Critical points}

 The fixed points of the system are obtained by setting $dx_1/dN=0$, $dx_2/dN=0$, $dx_3/dN=0$, $dy/dN=0$, $dz/dN$ and $d\lambda/dN=0$ in Eq. (\ref{dynsystemS})--(\ref{eq:dlambda/dnS}). When $\Gamma=1$, $\lambda$ is constant the potential is $V(\phi)=V_0e^{-\lambda \phi}$ \cite{copeland1998,ng2001}\footnote{The equation for $\lambda$ is also equal zero when $x_1=0$ or $\lambda=0$, so that  $\lambda$ should not necessarily be constant, for the fixed points with this value of $x_1$. However, for the case of dynamical $\lambda$, the correspondent eigenvalue is equal zero, indicating that the  fixed points is not hyperbolic.}. The fixed points are shown in Table \ref{criticalpointsS}. Notice that $x_3$ and $y$ cannot be negative and recall that $\Omega_r=z^2$. Some of the fixed points do not exist for the phantom field because for those cases $\Omega_\phi$ is negative.

 \begin{table*}\centering
\begin{tabular}{llllllllll}
\hline\noalign{\smallskip}
Point  & Existence & $x_1$ &$x_2$&$x_3$& $y$  &$z$& $w_\phi$ & $\Omega_\phi$& $w_{eff}$ \\\\
\noalign{\smallskip}\hline\noalign{\smallskip}
(a)   	&$ Q=0$	       & 0 & 0& any&0     & 0& --&0&0   \\
						
		(b) &$ \epsilon=+1$	 &$\frac{-\sqrt{6}Q}{3}$ & 0  & 0   & 0& 0& 1& $\frac{2Q^2}{3}$&  $\frac{2Q^2}{3}$ \\
		(c) & any & $0$&$0$ & any &0&1 &-- &0 & $\frac{1}{3}$ \\
		 (d) &$ \epsilon=+1$ & $\frac{-1}{\sqrt{6}Q}$ &0   &0& $0$ &$ \sqrt{1-\frac{1}{2Q^2}}$& $1$&$\frac{1}{6Q^2}$  &   $\frac{1}{3}$ \\ 
		 (e) &$ \epsilon=+1$ & $\frac{2\sqrt{6}}{3\lambda}$ & 0 &0 &$\frac{2\sqrt{3}}{3\lambda}$&$\sqrt{1-\frac{4}{\lambda^2}}$   &$\frac{1}{3}$  &$\frac{4}{\lambda^2}$ & $\frac{1}{3}$\\
		 
    (f) &$ \epsilon=+1$ & any&$\sqrt{1-x_1^2}$   &0& 0 & 0& 1&1  &   1\\ 
 
      (g) &any &$\frac{\sqrt{6}}{2(\lambda+Q)}$ & 0  & 0   & $\sqrt{\frac{2Q(Q+\lambda)+3\epsilon}{2(\lambda+Q)^2}}$ & $0$& $\frac{-Q(Q+\lambda)}{Q(Q+\lambda)+3\epsilon}$& $\frac{Q(Q+\lambda)+3\epsilon}{(\lambda+Q)^2}$& $\frac{-Q}{\lambda+Q}$\\
 (h) & any & $\frac{\epsilon \lambda}{\sqrt{6}}$ &0&0 & $\sqrt{1-\frac{\epsilon \lambda^2}{6}}$ & 0& $-1+\frac{\epsilon \lambda^2}{3}$ &1  & $-1+\frac{\epsilon \lambda^2}{3}$\\

 \noalign{\smallskip}\hline
\end{tabular}
\caption{\label{criticalpointsS} Critical points ($x_1$, $x_2$, $x_3$, $y$ and $z$) of the Eq. (\ref{dynsystemS}) for the quintessence and phantom field. It is shown the condition of existence, if any,  of the fixed point (point (a) exists only for $Q=0$, for instance, while the point (b) does not exist for the phantom field.).  The table shows the correspondent equation of state for the dark energy (\ref{eq:equationStateXYS}), the effective equation of state (\ref{eq:weffS}) and the density parameter for dark energy (\ref{eq:densityparameterXYS}).}
\end{table*}

The eingenvalues of the Jacobian matrix were found for each fixed point in Table \ref{criticalpointsS}.  The results are shown in Table \ref{stabilityS}.

\begin{table}
\centering
\begin{tabular}{llll}
\hline\noalign{\smallskip}
  Point & $\mu_1$ & $\mu_2$ & $\mu_3$ \\
\noalign{\smallskip}\hline\noalign{\smallskip}
  (a)        & $\frac{9}{2}$&$-\frac{3}{2}$& 0\\
  (b)    & $Q^2-\frac{3}{2}$&$Q^2-\frac{3}{2}$&$0$\\
  (c)  & $-1$&$-1$  &$0$\\ 
  (d)   & $-1+\frac{1}{2Q^2}$&$-1$   &   $0$      \\ 
   (e)   & & see the main text    & \\
   (f)   & $3x_1^2+\sqrt{6}Qx_1$& $3(x_1^2-1)$ &$\mp x_1\sqrt{1-x_1^2}$   \\
       (g)   & $-\frac{\lambda+4Q}{2(\lambda+Q)} $&$-\frac{3(\lambda+2Q)}{2(\lambda+Q)}$&0\\
  (h) & $-3+\epsilon \lambda(\lambda+Q) $&   $-3+\frac{\epsilon\lambda^2}{2} $&0 \\
  \hline
    &  $\mu_4$& $\mu_5$& Stability\\
\hline
  (a)       & $\frac{3}{2}$&$-\frac{1}{2}$&saddle\\
  (b)    & $Q(Q+\lambda)+\frac{3}{2}$& $Q^2+\frac{1}{2}$&unstable  or saddle\\
  (c)  & 2& $1$&saddle\\ 
   
  (d)       &$2+\frac{\lambda}{2Q}$ & $1-\frac{Q^2}{2}$&saddle\\ 
  (e)  & & see the main text    & \\
   (f)   &$3-\frac{\sqrt{6}x_1\lambda}{2}$& $1$ &unstable or saddle \\
        (g)  & $\mu_{4d}$& $\mu_{5d}$&saddle or stable\\
  (h)   &  $-3+\frac{\epsilon\lambda^2}{2}  $&$-2+\frac{\epsilon\lambda^2}{2}  $ &saddle or stable\\
    
     \noalign{\smallskip}\hline
\end{tabular}
\caption{\label{stabilityS} Eigenvalues and stability of the fixed points for the quintessence (phantom) field.}
\end{table}

The eigenvalues $\mu_{4e}$ and $\mu_{5e}$ are

\begin{eqnarray}\label{eigenh}
&\mu_{4e,5e}=-\frac{3(\lambda+2Q)}{4(\lambda+Q)}\times\nonumber\\&\left(1\pm\sqrt{1+\frac{8[3-\epsilon\lambda(\lambda+Q)][3\epsilon+2Q(\lambda+Q)]}{3(\lambda+2Q)^2}}\right).
\end{eqnarray}

At the first sight, one might think that  the linear analysis would not give a complete description of the stability, because all fixed points, but (f), have at least one eigenvalue equals zero. However, as pointed out in \cite{Boehmer:2011tp}, fixed points that have at least one positive and one negative eigenvalue are always unstable, and methods such as center manifold \cite{Boehmer:2011tp} should be used to analyze the stability of the critical points  that can be stable [((g) and (h)]. Even so, for almost all fixed points, but (a) and (c),  $x_3=0$, which means $\phi\rightarrow \infty$. However, this limit implies that $x_2\propto  \phi\dot{\theta}/H\rightarrow \infty$ as well,  provided that $H$ is finite. This issue occured for the points (b) and (d) to (h), as can be seen in the Table \ref{criticalpointsS} whose mathematical inconsistency indicates that the these critical points are  not physically acceptable.

Since all critical points are similar to those ones found in the literature \cite{copeland1998,Amendola:1999er,Gumjudpai:2005ry} we reproduce the main results in the appendix \ref{appB} for the sake of completeness, which are valid for the case of real scalar field.

\subsection{Summary}

From the eight fixed points presented in the quintessence (phantom) case, only (a) and (c) are physically viable and they describe  the sequence: radiation $\rightarrow$ matter. Both of them are unstable, however there does not exist a point that describe the dark-energy-dominated universe. Thus, the  extra degree of freedom due to the phase $\theta$ spoils the physically acceptable fixed points that exist for the case of real scalar field, indicating that the dynamical system theory is not a good tool when one tries to analyze the complex quintessence (phantom).




\section{Tachyon dynamics}\label{tachdyn}

 The complex tachyon field $\Phi=\phi e^{i\theta}$, where $\phi$ is the absolute value of the tachyon field and $\theta$ is a phase, is described by the Born-Infeld Lagrangian
 
\begin{equation}\label{LBI}
 \mathcal{L}_{BI}=-\sqrt{-g}V(|\Phi|)\sqrt{1-\partial^\mu\Phi\partial_\mu\Phi},
\end{equation} 

\noindent where $V(\phi)$ is the tachyon potential, which depends only on the absolute value of the scalar field $\phi\equiv|\Phi|$.  For a homogeneous field $\phi\equiv\phi(t)$ and $\theta\equiv\theta(t)$, in an expanding universe with Friedmann-Robertson-Walker metric, the Lagrangian becomes

\begin{equation}\label{LBIcosmo}
 \mathcal{L}_{BI}=-a^3V(\phi)\sqrt{1-\dot{\phi}^2-\phi^2\dot{\theta}^2},
\end{equation}

\noindent where $a\equiv a(t)$ is the scale factor. The equations of motion for $\phi$ and $\theta$ are respectively

\begin{eqnarray}\label{eqmotion}
 \frac{\ddot{\phi}}{1-\phi^2\dot{\theta}^2-\dot{\phi}^2}+&\frac{\ddot{\theta}+\dot{\phi}\dot{\theta}\phi^{-1}}{(1-\phi^2\dot{\theta}^2-\dot{\phi}^2)}\frac{\phi^2\dot{\phi}\dot{\theta}}{(1-\phi^2\dot{\theta}^2)}\nonumber\\
 &+\frac{3H\dot{\phi}-\phi\dot{\theta}^2}{1-\phi^2\dot{\theta}^2}+\frac{V'(\phi)}{V(\phi)}=0,
\end{eqnarray}

\begin{equation}\label{eqmotion2}
 \frac{\ddot{\theta}}{1-\phi^2\dot{\theta}^2-\dot{\phi}^2}+3H\dot{\theta}+\frac{2\dot{\phi}\dot{\theta}}{\phi(1-\phi^2\dot{\theta}^2-\dot{\phi}^2)}=0,
\end{equation}

\noindent where the prime denotes time derivative with respect to $\phi$. When the phase $\theta$ is zero, we recover the well-known equation of motion for the tachyon field.

The Friedmann equations for the complex tachyon, in the presence of barothorpic fluids,  are

\begin{equation}\label{eq:1stFEmatter}
  H^2=\frac{1}{3}\left(\frac{V(\phi)}{\sqrt{1-\phi^2\dot{\theta}^2-\dot{\phi}^2}}+ \rho_m+\rho_r\right),
\end{equation}

\begin{equation}\label{eq:2ndFEmatter}
  \dot{H}=-\frac{1}{2}\left(\frac{V(\phi)(\dot{\phi}^2+\phi^2\dot{\theta}^2)}{\sqrt{1-\phi^2\dot{\theta}^2-\dot{\phi}^2}} +\rho_m+\frac{4}{3}\rho_r\right),
\end{equation}

The equation of state for the tachyon field yields

\begin{equation}\label{eqstate}
 w_\phi=\frac{p_\phi}{\rho_\phi}=\dot{\phi}^2+\phi^2\dot{\theta}^2-1,
\end{equation}

\noindent thus, the tachyon behavior is between the cosmological constant one ($w_\phi=-1$) and matter one ($w_\phi=0$). 

The interaction between the tachyon and the dark matter is driven by the coupling $\mathcal{Q}=Q \rho_m\rho_\phi\dot{\phi}/H$, which in turn we consider that it modifies the right-hand side of Eq. (\ref{eqmotion}). With this form of coupling, the time dependence of the coupling is implicit in the Hubble parameter $H$. We are now ready to proceed the dynamical analysis of the system.

\subsection{Autonomous system}\label{autosystem}

The dimensionless variables for the case of tachyon field are

\begin{eqnarray}\label{eq:dimensionlessXY}
 x_1\equiv & \dot{\phi}, \quad x_2\equiv \phi\dot{\theta}, \quad x_3\equiv \frac{1}{H\phi}, \quad y\equiv \frac{\sqrt{V(\phi)}}{\sqrt{3}H}, 
\nonumber\\
 & z\equiv\frac{\sqrt{\rho_r}}{\sqrt{3}H}, \quad \lambda\equiv -\frac{V'}{V^{3/2}}, \quad \Gamma\equiv \frac{VV''}{V'^2}.
\end{eqnarray}

\noindent Since $\dot{\phi}$ and $\theta$ are dimensionless variables, $\phi$ has dimension of time. 

The dark energy density parameter is written in terms of these new variables as

\begin{equation}\label{eq:densityparameterXY}
 \Omega_\phi \equiv \frac{\rho_\phi}{3H^2} = \frac{y^2}{\sqrt{1-x_1^2-x_2^2}},
 \end{equation}

\noindent so that Eq. (\ref{eq:1stFEmatter}) can be written as 

\begin{equation}\label{eq:SomaOmegas}
\Omega_\phi+\Omega_m+\Omega_r=1,
\end{equation}

\noindent where the matter and radiation density parameter are defined by $\Omega_i=\rho_i/(3H^2)$, with $i=m,r$. From Eqs. (\ref{eq:densityparameterXY}) and (\ref{eq:SomaOmegas}) we have that $x_1$, $x_2$ and $y$ are restricted in the phase plane by the relation

\begin{equation}\label{restriction}
0\leq x_1^2+x_2^2+y^4\leq 1,
 \end{equation}
 
\noindent due to $0\leq \Omega_\phi\leq 1$. In terms of these new variables the equation of state $w_\phi$  is

\begin{equation}\label{eq:equationStateXY}
 w_\phi =x_1^2+x_2^2-1,
\end{equation}

\noindent which is clearly a trivial extension for the complex scalar field. The total effective equation of state is

\begin{equation}\label{eq:weff}
 w_{eff} = \frac{p_\phi+p_r}{\rho_\phi+\rho_m+\rho_r}=-y^2\sqrt{1-x_1^2-x_2^2}+\frac{z^2}{3},
\end{equation}

\noindent with an accelerated expansion for  $w_{eff} < -1/3$. The dynamical system for the variables  $x_1$, $x_2$, $x_3$, $y$, $z$  and $\lambda$ are

\begin{eqnarray}\label{dynsystem}\label{eq:dx1/dn}
\frac{dx_1}{dN}&=-(1-x_1^2-x_2^2)\times \nonumber\\&\left[3x_1-\sqrt{3}y\lambda+3Q\left(1-z^2-\frac{y^2}{\sqrt{1-x_1^2-x_2^2}}\right)\right]+x_2^2x_3,\end{eqnarray}

\begin{equation}\label{eq:dx2/dn}
\frac{dx_2}{dN}=-x_1x_2x_3-3x_2(1-x_1^2-x_2^2),\end{equation}

\begin{equation}\label{eq:dx3/dn}
\frac{dx_3}{dN}=-x_1x_3^2+\frac{x_3}{2}\left[3+z^2-\frac{3y^2(1-x_1^2-x_2^2)}{\sqrt{1-x_1^2-x_2^2}}\right],\end{equation}

\begin{equation}\label{eq:dy/dn}
\frac{dy}{dN}=\frac{y}{2}\left[-\sqrt{3}x_1y\lambda+3+z^2- \frac{3y^2(1-x_1^2-x_2^2)}{\sqrt{1-x_1^2-x_2^2}}\right],
\end{equation}

\begin{equation}\label{eq:dz/dn}
\frac{dz}{dN}=-2z+\frac{z}{2}\left[3+z^2-\frac{3y^2(1-x_1^2-x_2^2)}{\sqrt{1-x_1^2-x_2^2}}\right],
\end{equation}

\begin{equation}\label{eq:dlambda/dn}
\frac{d\lambda}{dN}=-\sqrt{3}\lambda x_1y\left(\Gamma-\frac{3}{2}\right).
\end{equation}

\subsection{Critical points}

 The fixed points of the system are obtained by setting $dx_1/dN=0$, $dx_2/dN=0$, $dx_3/dN=0$, $dy/dN=0$, $dz/dN$ and $d\lambda/dN=0$ in Eq. (\ref{dynsystem}). When $\Gamma=3/2$, $\lambda$ is constant the potential has the form found in refs.~\cite{Aguirregabiria:2004xd,Copeland:2004hq} ($V(\phi)\propto \phi^{-2}$), known in the literature for both coupled \cite{Gumjudpai:2005ry,micheletti2009} and uncoupled \cite{Padmanabhan:2002cp,Bagla:2002yn} dark energy\footnote{The equation for $\lambda$ is also equal zero when $x_1=0$, $y=0$ or $\lambda=0$, so that  $\lambda$ should not necessarily be constant, for the fixed points with these values of $x_1$ or $y$. However, for the case of dynamical $\lambda$, the correspondent eigenvalue is equal zero, indicating that the  fixed points is not hyperbolic.}.  The fixed points are shown in Table \ref{criticalpoints}. Notice that $x_3$ and $y$ cannot be negative and recall that $\Omega_r=z^2$.

 \begin{table*}
 \centering
\begin{tabular}{llllllllll}
\hline\noalign{\smallskip}
Point & $x_1$ &$x_2$&$x_3$& $y$ & $z$& $w_\phi$ & $\Omega_\phi$& $w_{eff}$ \\
\noalign{\smallskip}\hline\noalign{\smallskip}
(a1)         & $\pm 1$ & 0& 0&0& 0& 0&0&0   \\
(a2)         & $0$ &$\pm 1$& 0& 0&0& 0& 0&0   \\
  (a3) 						& 1 &0&$ \frac{3}{2}$& 0&0& 0& 0&0  \\
   (a4) & $-Q$&$\pm\sqrt{1-Q^2}$ & 0 &0 &0 &0 &0  & 0\\
  (b) &any &$\pm\sqrt{1-x_1^2}$  & 0   & 0& $\pm 1$& 0& 0&  $\frac{1}{3}$  \\
     (c) &1 & 0  & 2   & 0& $\pm 1$& $0$& 0&  $\frac{1}{3}$  \\
      (d) &0 & 0  & any   & 0& $\pm 1$& $-1$& 0&  $\frac{1}{3}$ \\
           (e) & 0 &0&any & 1 & 0& $-1$ &1  & $-1$\\
  (f1) & $\frac{\lambda y_c}{\sqrt{3}}$ &0   &0& $y_c$ & 0& $\frac{\lambda^2 y_c^2}{3}-1$&1  &   $w_\phi$\\ 
  (f2) & $\frac{\lambda y_c}{\sqrt{3}}$ &0   &$\frac{\sqrt{3}\lambda y_c}{2}$& $y_c$ & 0& $\frac{\lambda^2 y_c^2}{3}-1$&1  &   $w_\phi$\\ 
  (g) & $-Q$ & 0 &0 &0&0   &$Q^2-1$  &0  & 0\\
   (h1) & $x_f$ &0&  0& $y_f$ &0 &$x_f^2-1$ &$\frac{w_{eff}}{w_\phi}$  &$\frac{x_fy_f\lambda}{\sqrt{3}}-1$ \\ 
 (h2) & $x_f$ &0&  $\frac{\sqrt{3\lambda y_f}}{2}$& $y_f$ &0 &$x_f^2-1$ &$\frac{w_{eff}}{w_\phi}$  &$\frac{x_fy_f\lambda}{\sqrt{3}}-1$ \\ 
  \noalign{\smallskip}\hline
\end{tabular}
\caption{\label{criticalpoints} Critical points ($x_1$, $x_2$, $x_3$, $y$ and $z$) of the Eq. (\ref{dynsystem}), for the tachyon field. The table shows the correspondent equation of state for the dark energy (\ref{eq:equationStateXY}), the effective equation of state (\ref{eq:weff}) and the density parameter for dark energy (\ref{eq:densityparameterXY}).}
\end{table*}

The fixed points $y_c$, $x_f$ and $y_f$ are shown below

\begin{equation}\label{yc}
y_c=\sqrt{\frac{\sqrt{\lambda^4+36}-\lambda^2}{6}},
\end{equation}

\begin{equation}\label{xf}
x_f=-\frac{Q}{2}\pm \frac{\sqrt{Q^2+4}}{2},
\end{equation}

\begin{equation}\label{yf}
y_f=\frac{-\lambda x_f+ \sqrt{\lambda^2x_f^2+12\sqrt{1-x_f^2}}}{\sqrt{12(1-x_f^2)}}.
\end{equation}

The eingenvalues of the Jacobian matrix were found for each fixed point in Table \ref{criticalpoints}.  The results are shown in Table \ref{stability}.

\begin{table*}
\begin{tabular}{lllllll}
\hline\noalign{\smallskip}
  Point & $\mu_1$ & $\mu_2$ & $\mu_3$ & $\mu_4$& $\mu_5$& Stability\\
\noalign{\smallskip}\hline\noalign{\smallskip}
  (a1)        & $6(1\pm Q)$&0& $\frac{3}{2}$& $\frac{3}{2}$&$-\frac{1}{2}$&saddle\\
  (a2)        & $0$&  $6$&$\frac{3}{2}$ & $\frac{3}{2}$&$-\frac{1}{2}$&saddle\\
   (a3)        & $6(1+ Q)$&$-\frac{3}{2}$& $-\frac{3}{2}$ & $\frac{3}{2}$&$-\frac{1}{2}$&saddle\\
  (a4)  & $0$&$6(1-Q^2)$  &$\frac{3}{2}$& $\frac{3}{2}$& $-\frac{1}{2}$&saddle\\ 
  (b)    & $6x^2_1$&$6x_2^2$&2 & 2& 1 &unstable\\
  (c)    & $6$&$-2$&$-2$ & 2& 1&saddle\\
   (d)   & $-3 $&$-3$&2& 4& 1&saddle\\
  (e)   & $-3 $&   $-3 $&0 &  $-3 $&$-2 $ &stable \\
 (f1)   & $\sqrt{3}Q\lambda y_c -3\left(1-\frac{\lambda^2y_c^2}{3}\right)$&$-3\left(1-\frac{\lambda^2y_c^2}{3}\right)$   &   $\frac{\lambda^2y_c^2}{2}$   &$\frac{\lambda^2y_c^2}{2}-3$ & $\frac{\lambda^2y_c^2}{2}-2$&saddle    \\ 
 (f2)   & $\sqrt{3}Q\lambda y_c -3\left(1-\frac{\lambda^2y_c^2}{3}\right)$&$\frac{\lambda^2y_c^2}{2}-3$   &   $-\frac{\lambda^2y_c^2}{2}$      &$\frac{\lambda^2y_c^2}{2}-3$ & $\frac{\lambda^2y_c^2}{2}-2$&stable for $\lambda<0$ or $Q=0$\\ 
  (g)  & $-3(1-Q^2)$&$-3(1-Q^2)$  &$\frac{3}{2}$& $\frac{3}{2}$& $-\frac{1}{2}$&saddle\\ 
    (h1)   & $3\left(x_f^2-\frac{x_fy_f \lambda}{\sqrt{3}}\right)$& $-3(1-x_f)$ &$\frac{3}{2}$   &$\frac{3}{2}\left(\frac{x_fy_f \lambda}{\sqrt{3}}-2\right)$& $\frac{3}{2}\left(\frac{x_fy_f \lambda}{\sqrt{3}}-\frac{4}{3}\right)$ &saddle  \\
    (h2)   & $3\left(x_f^2-\frac{x_fy_f \lambda}{\sqrt{3}}\right)$& $-\frac{\sqrt{3}}{2}\lambda x_fy_f-3(1-x_f)$ &$-\frac{\sqrt{3}}{2}\lambda x_f y_f$ &$\frac{3}{2}\left(\frac{x_fy_f \lambda}{\sqrt{3}}-2\right)$& $\frac{3}{2}\left(\frac{x_fy_f \lambda}{\sqrt{3}}-\frac{4}{3}\right)$  &stable \\
  
\noalign{\smallskip}\hline
\end{tabular}
\caption{\label{stability} Eigenvalues and stability of the fixed points for the tachyon field.}
\end{table*}

The points (a1)--(a4) correspond to a matter-dominated solution, since $\Omega_m=1$ and $w_{eff}=0$.  They are saddle points because at least one eigenvalue has an opposite sign. The point (a4) is actually the point (a1), with $Q=1$. Points (b), (c) and (d) are radiation-dominated solutions, with $\Omega_r=1$ and $w_{eff}=1/3$. The difference between them is that (b) and (c) have $w_\phi=0$, while (d) has $w_\phi=-1$ and admits any value for $x_3$. They are unstable [(b)] or saddle points [(c) or (d)]. 

The point (e) is in principle a dark-energy-dominated solution with $\Omega_\phi=1$ and $w_{eff}=w_\phi=-1$, whose existence is restrict to $\lambda=0$. However a careful analysis shows that the Jacobian matrix for this critical point has  zero eigenvector, thus it cannot be considered.  Points (f1) and (f2) are also a dark-energy-dominated solution ($\Omega_\phi=1$) whose equation of state depends on $\lambda$, which in turn can be either constant or zero. The case with constant $\lambda$ are shown in the Table \ref{criticalpoints} and an accelerated expansion occurs for $\lambda^2<2/\sqrt{3}$.  For $\lambda=0$ we recover the point (e). The  eigenvalues $\mu_2$, $\mu_4$ and $\mu_5$ of the fixed point (f1) and (f2) are always negative. For these points $\frac{\lambda^2y_c^2}{3}\leq 1$, then the first eigenvalue is also negative  if $Q=0$,  $\lambda<0$ or $\sqrt{3}Q\lambda y_c<3$. Therefore, the point (f2) describes a dark-energy-dominated universe and can lead to a late-time accelerated universe  if the requirement  $\mu_1<0$ is statisfied. On the other hand, (f1) is a saddle point. The effective equation of state depends only on $\lambda$, so the coupling $Q$ only changes the property of the fixed point.  

 The point (g) is also a saddle point with a matter-dominated solution, however, different from (a1)--(a4), the equation of state for the dark energy $w_\phi$ is no longer zero, but depends on $Q$, leading to an universe with accelerated expansion for $Q^2<2/3$. For this point the coupling is restrict to values $0\leq Q^2\leq 1$.

The last fixed points (h1)  and (h2) are valid for $x_f\neq 0$\footnote{The case for $x_f=0$ is the fixed point (e).}, for $Q\neq 0$ and for constant $\lambda$, and its behavior depends on $Q$. In order to have $x^2_f\leq1$, we must have $Q>0$ for the case with plus sign in $x_f$ (\ref{xf}), while we have $Q<0$ for the minus sign case. We restrict our attention for the plus sign case. When $Q\rightarrow \infty$, $x_f\rightarrow 0$ and $y_f\rightarrow 1$, in agreement with the restriction (\ref{restriction}). In addtition, as pointed out in ref.~\cite{Landim:2015poa}, the fixed points exists for some values of $\lambda>0$ and $Q$, due to Eq. (\ref{restriction}). Both fixed points have similar behaviour, however, (h1) is a saddle point, while (h2) is stable. Such difference is due to the eigenvalue $\mu_3$ (Table \ref{stability}). The eigenvalues $\mu_4$ and $\mu_5$ are always negative because $w_{eff}$ is between zero and minus one. The first eigenvalue is also negative because $\Omega_\phi\leq 1$, thus $x_f^2-1\leq\frac{x_fy_f \lambda}{\sqrt{3}}-1$, therefore $\frac{x_fy_f\lambda}{\sqrt{3}}\geq x_f^2$, since $x_f$ is always positive. Therefore, the point (h2) can lead to a late-time accelerated universe, depending on the value of $\lambda$ and $Q$.

As in the case of quintessence and phantom, the fixed points that have $x_3=0$ [(a1), (a2), (a4), (b), (f1), (g) and (h1)] indicate that  $\phi\rightarrow \infty$ and therefore  $x_2\equiv  \phi\dot{\theta}\rightarrow \infty$ as well. However this limit is in contradiction to what is presented in Table \ref{criticalpoints} for $x_2$ showing that these seven critical points are not physically acceptable.

All fixed points reproduce the previous results in the literature \cite{Gumjudpai:2005ry,Aguirregabiria:2004xd,Copeland:2004hq,Landim:2015poa} and they are generalizations of those analyses, with same stability behaviour for the critical points. This indicates that the degree of freedom due to the complex scalar has no effect on the stability and on  the evolution of the system of equations, when compared with the case of real scalar field.

\subsection{Summary}

The critical points showed in the tachyonic case describe the three phases of the universe: the radiation-dominated era, the matter-dominated era, and the present dark-energy-dominated universe. The matter-dominated universe can be described by the saddle point (a3). There are two points that  can represent the radiation-dominated era: (c) and (d). The two points are saddle, with the additional difference that the point (d) has an  equation of state for dark energy equals to minus one. 

A tachyonic-dominated universe is described by  the point point  (f2) and (h2).  The point (f2) can be stable only if the coupling is zero or $\lambda<0$. The last fixed point (h2)  is stable and can describe an accelerated universe depending on the value of $\lambda$ and $Q$. 


From all the critical points,  the cosmological transition radiation $\rightarrow$ matter $\rightarrow$ dark energy is achieved considering the following sequence of fixed points:  (c) or (d) $\rightarrow$ (a3) $\rightarrow$ (f2) [$\lambda$ dependent] or (h2) [$Q$ and $\lambda$ dependent]. Although the sequence is viable, the form of the potential dictates whether the fixed points are allowed or not.  Among several possibilities in the literature, the potential $V(\phi)\propto \phi^{-n}$, for instance, leads to a dynamically changing $\lambda$ (either if $\lambda\rightarrow 0$ for $0<n<2$, or $\lambda\rightarrow \infty$ for  $n>2$) \cite{Abramo:2003cp}. A dynamically changing $\lambda$ is allowed for the fixed points (a3), (c) and (d). On the other hand, points (f2) and (h2) require a constant $\lambda$, implying $V(\phi)\propto \phi^{-2}$ \cite{Gumjudpai:2005ry,micheletti2009,Padmanabhan:2002cp,Bagla:2002yn}.

 \section{Conclusions}\label{conclu}
 
 In this paper we studied coupled dark energy using a complex scalar field, in the light of the dynamical system theory. There were analyzed three possibilities: quintessence, phantom and tachyon field.  All three possibilities are known  in the literature for the real field \cite{Amendola:1999er,Gumjudpai:2005ry,ChenPhantom,Landim:2015poa}, and for uncoupled and complex quintessence field \cite{Zhai2005}. Thus, a natural question that arises is  how a complex scalar field changes the previous results  and if there are new fixed points due to the complex field. Although some equations for the dimensionless variables are trivial extensions of the real field case (e.g. the equation of state for the scalar field), the differential equations were generalized. All fixed points found here are in agreement with the previous results, with no new fixed points, however there are some crucial differences. For the quintessence and the phantom there is a contradiction between the fixed points $x_2$ and $x_3$ when the latter is zero.  This situation occurs for almost all fixed points and the only two exceptions are unstable points that  represent respectively the radiation and matter era, so the dark-energy-dominated universe is absent. Therefore the extra degree of freedom spoils the results known in the case of real scalar field. For the tachyon field all the critical points are also similar to the real field case, with same stabilities.  Therefore, the extra degree of freedom due to the complex tachyon field plays no role on the stability of the critical points. Although the results presented here enlarge the previous results found in the literature, with the generalization of the equations of motion,  the dynamical system theory does not provide further information in what is already known for the case of real scalar field, letting open the possibility of studing complex scalar fields by other ways of analysis.


\section{Appendix}\label{appB}

In this appendix we reproduce the results in the literature \cite{copeland1998,Amendola:1999er,Gumjudpai:2005ry} regarding the fixed points presented in Table \ref{criticalpointsS}, for the sake of completeness.

The fixed point (a) is a saddle point which describes a matter-dominated universe, however it is valid only for $Q=0$. The other possibility of matter-dominated universe with $Q\neq 0$ arises from the fixed point (b). This point is called ``$\phi$-matter-dominated epoch'' ($\phi$MDE) \cite{Amendola:1999er} and it can be either unstable or a saddle point. However, due to $\Omega_\phi=2Q^2/3\ll 1$, the condition $Q^2\ll 1$ should hold in order to the point be responsible for the matter era. Thus, $\mu_1$ and $\mu_2$ are negative, while $\mu_5$ is always positive and $\mu_4$ is positive for $Q(\lambda+Q)>-3/2$. Therefore (b) is a saddle point. 

The radiation-dominated universe is described by the critical points (c), (d) and (e), only for the quintessence field. The first two points are saddle, as it is easily seen in table  \ref{stabilityS}, and the last one had its stability described numerically in \cite{Amendola:1999er}. However, both (d) and (e) are not suitable to describe the universe we live in, due to nucleosynthesis constraints \cite{Amendola,amendola2001}. The nucleosynthesis bound $\Omega_\phi^{BBN} < 0.045$ \cite{bean2001} implies $Q^2>3.7$ for the point (d) and $\lambda^2>88.9$ for the point (e). Thus, the requirement for the point (d) is not consistent with the condition of point (b) and the constraint on $\lambda^2$ does not allow a scalar field attractor, as we will se soon. Therefore, the only viable cosmological critical point for the radiation era is (c).

The point (f) is an unstable or saddle point and it does not describe an accelerated universe.  The last possibility for the matter era is the point (g), with eingenvalues showed in table \ref{stabilityS} and Eq. (\ref{eigenh}). Since $w_{eff}\simeq 0$ for $|\lambda|\gg |Q|$, the fixed point is either stable or stable spiral, hence the universe would not exit from the matter dominance. 

On the other hand, the point (g) can lead to an accelerated universe, for the quintessence field case ($\epsilon=+1$), provided that $3<\lambda(\lambda+Q)$, because $\Omega_\phi\leq 1$, and $Q>\lambda/2$, from $w_{eff}<-1/3$. Regarding $\lambda>0$, the two eigenvalues $\mu_1$ and $\mu_2$ are always negative and since $Q>3/\lambda-\lambda$, the behaviour of $\mu_{4d,5d}$ depends on the second term in the square root of (\ref{eigenh})

\begin{equation}
A\equiv \frac{8[3-\lambda(\lambda+Q)][3\epsilon+2Q(\lambda+Q)]}{3(\lambda+2Q)^2}.\end{equation}

 \noindent From the condition $3<\lambda(\lambda+Q)$ we have $A<0$, and if $A<1$ the fixed point is stable. Otherwise, i.e. $A>1$, the critical point is a stable spiral. Thus, the value of the coupling dictates which behaviour the fixed point will have: stable for $3/\lambda-\lambda<Q<Q_*$ or stable spiral for $Q>Q_*$, where $Q_*$ is the solution of $A=1$. However, even in the case where one can get $\Omega_\phi\simeq 0.7$ \cite{Hebecker2000,amendola2001}, there are no allowed region in the $(Q,\lambda)$ plane corresponding to the transition from $\phi $MDE to scaling attractor \cite{Amendola:1999er}. Thus, it is hard to gather the conditions for the point $\phi$MDE and the point (g). For the case of the phantom field ($\epsilon=-1$), the condition $y^2>0$ implies   $2Q(Q+\lambda)>3$. Hence, $\mu_4<0$ and $\mu_5>0$, and (g) is a saddle point.
 
 The last fixed point (h) leads to an accelerated universe provided that $\lambda^2<2$. With this condition, the eingenvalues $\mu_2$, $\mu_4$ and $\mu_5$ are always negative. The first eigenvalue $\mu_1$ is also always negative for the phantom field, and it is for the quintessence field with the condition $\lambda(\lambda+Q)<3$. Therefore, the point is stable if the previous conditions are satisfied.

\begin{acknowledgements}
I thank Elcio Abdalla and Giancarlo Camilo for various suggestions and comments, during all the steps of the work. I also thank an anonymous reviewer for his essential comments. This work is supported by FAPESP Grant No. 2013/10242-1. \end{acknowledgements}

\bibliographystyle{unsrt}
\bibliography{trab1}\end{document}